\documentclass[prd,showpacs,preprintnumbers,superscriptaddress]{revtex4}
\usepackage{amssymb,easybmat,graphicx}

\newcommand{\bvec}[1]{\ensuremath{\mbox{\boldmath $\mathrm{#1}$}}}

\begin{document}

\title{Excited-State Effective Masses in Lattice QCD}

\author{George T.\ Fleming}
\email{George.Fleming@Yale.edu}
\affiliation{Sloane Physics Laboratory, Yale University, New Haven, CT
  06520, USA}

\author{Saul D.\ Cohen}
\email{sdcohen@jlab.org}
\affiliation{Thomas Jefferson National Accelerator Facility, Newport News,
  VA 23606, USA}

\author{Huey-Wen Lin}
\email{hwlin@jlab.org}
\affiliation{Thomas Jefferson National Accelerator Facility, Newport News,
  VA 23606, USA}

\author{Victor Pereyra}
\email{victor@wai.com}
\affiliation{Weidlinger Associates Inc., Mountain View, CA 94040, USA}

\begin{abstract}

We apply black-box methods, \textit{i.e.}\ where the performance of the method
does not depend upon initial guesses, to extract excited-state energies from
Euclidean-time hadron correlation functions.  In particular, we extend the
widely used effective-mass method to incorporate multiple correlation
functions and produce effective mass estimates for multiple excited states.
In general, these excited-state effective masses will be determined by finding
the roots of some polynomial.  We demonstrate the method using sample lattice
data to determine excited-state energies of the nucleon and compare the
results to other energy-level finding techniques.

\end{abstract}


\maketitle 

\section{\label{sec:introduction}Introduction}

Lattice quantum chromodynamics (LQCD) has been used successfully to compute
many experimentally observable quantities from first-principles calculation of
Euclidean-time hadron correlation functions, even occasionally predicting
experimental results before they are measured.  However, the successes of LQCD
have mostly been restricted to computation of the physical properties of the
lowest-energy states in each quantum number
channel by focusing on the large-time behavior of correlation functions where
uncertainties due to excited-state contributions are exponentially suppressed.
Given that signal-to-noise in correlation functions also falls exponentially
at large times, success is often dictated by available computational
resources.

Both in meson and baryon spectroscopy there are many experimentally observed
excited states whose physical properties are poorly understood that could use
theoretical input from LQCD to solidify their identification.  Other
excited-state quantities that could be computed on the lattice, such as form
factors and coupling constants, would be useful to groups such as the Excited
Baryon Analysis Center (EBAC) at Jefferson Lab, where dynamical reaction
models have been developed to interpret experimentally observed properties of
excited nucleons in terms of QCD~\cite{Lee:2006xu,Matsuyama:2006rp}.  In
certain cases, input from the lattice may be helpful in determining the
composition of controversial states, which may be interpreted as ordinary
hadrons, tetra- or pentaquarks, hadronic molecules or unbound resonances.

Among the excited nucleon states, the nature of the Roper resonance, $N(1440)
\ P_{11}$, has been the subject of interest since its discovery in the 1960's.
It is quite surprising that the rest energy of the first excited state of the
nucleon is less than the ground-state energy of nucleon's negative-parity
partner, the $N(1535) S_{11}$ \cite{Yao:2006px}, a phenomenon never observed
in meson systems.  There are several interpretations of the Roper state, for
example, as the hybrid state that couples predominantly to QCD currents with
some gluonic contribution~\cite{Carlson:1991tg} or as a five-quark
(meson-baryon) state~\cite{Krehl:1999km}.

\begin{figure}
  \includegraphics[width=2.5in]{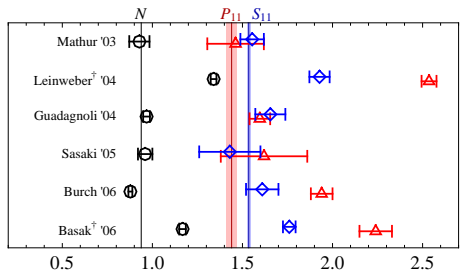}
  \caption{\label{fig:all-roper}Summary of previous lattice calculations with
    extrapolation to the physical pion mass point or the lowest simulated pion
    point (labeled as ``$\dag$'').}
\end{figure}

Early LQCD calculations using the quenched approximation \cite{Sasaki:2001nf,
Mathur:2003zf,Guadagnoli:2004wm,Leinweber:2004it,Sasaki:2005ap,Sasaki:2005ug,
Burch:2006cc}, found the computed spectrum inverted relative to experiment,
with $P_{11}$ heavier than the $S_{11}$.  A recent study \cite{Mathur:2003zf}
suggested that qualitative agreement between experiment and LQCD in the
quenched approximation could be restored provided other simulation effects due
to finite volumes and unphysically heavy quarks were properly addressed.  The
study strongly suggests the nature of the Roper resonance changes dramatically
as the quarks are made physically light in LQCD simulations, as in
Fig.~\ref{fig:all-roper}.  Clearly, future LQCD calculations will require
improved analysis techniques for extracting multiple excited-state energies,
as well as variational wavefunctions, in the nucleon sector to test the
validity of this claim.

\begin{table}
\begin{center}
\begin{tabular}{|c|c|c|c|c|c|c|c|}
\hline
Group & $N_{\rm f}$ & $S_{\rm f}$ & $a_t^{-1}$ (GeV)  & $M_\pi$ (GeV) & $L$ (fm) & Method &Extrapolation \\
\hline \hline
Basak et~al.~\cite{Basak:2006ww}    & 0   & Wilson & 6.05   & 0.49 &  2.35  & VM & N/A \\
Burch et~al.~\cite{Burch:2006cc}    & 0   & CIDO & 1.68,1.35   & 0.35--1.1 &  2.4  & VM & $a+b m_\pi^2$ \\
Sasaki et~al.~\cite{Sasaki:2005ap}    & 0   & Wilson & 2.1   & 0.61--1.22 &  1.5,3.0  & MEM & $\sqrt{a+b m_\pi^2}$ \\
Guadagnoli et~al.~\cite{Guadagnoli:2004wm}    & 0   & Clover~\cite{Sheikholeslami:1985ij} & 2.55   & 0.51--1.08 &  1.85  & SBBM & $a+b m_\pi^2 +c m_\pi^4$ \\
Leinweber et~al.~\cite{Leinweber:2004it}    & 0   & FLIC & 1.6   & 0.50--0.91 &  2.0  & VM  & N/A \\
Mathur et~al.~\cite{Mathur:2003zf}    & 0   & Overlap~\cite{Neuberger:1997fp} & 1.0   & 0.18--0.87 &  2.4,3.2  & CCF  & $a+b m_\pi +c m_\pi^2$ \\
Sasaki et~al.~\cite{Sasaki:2001nf}    & 0   & DWF & 2.1   & 0.56--1.43 &  1.5  & VM  & $a+b m_\pi^2$ \\
\hline
\end{tabular}
\end{center}
\caption{\label{tab:SP_summary}Summary of existing published $S_{11}$ and
  $P_{11}$ calculations. Due to space  limitations, we adopt these
  abbreviations for fermion actions: Domain-Wall
  Fermions~\cite{Kaplan:1992bt,Kaplan:1992sg,Shamir:1993zy,Furman:1994ky}
  (DWF), Chirally Improved Dirac
  Operator~\cite{Gattringer:2000js,Gattringer:2000qu} (CIDO), Fat-Link
  Irrelevant Clover~\cite{Zanotti:2001yb} (FLIC); and for the analysis
  methods: Variational Method~\cite{Michael:1985ne,Luscher:1990ck} (VM),
  Constrained Curve Fitting~\cite{Lepage:2001ym} (CCF), Maximum Entropy
  Method~\cite{Nakahara:1999vy,Asakawa:2000tr} (MEM), Simplified Black Box
  Method~\cite{Fleming:2004hs,Guadagnoli:2004wm} (SBBM). For those works which
  do not perform extrapolation, we use the lightest pion mass to represent
  their results.}
\end{table}

Apart from the vast amount of detail about excited states accessible to LQCD
computations with advanced analysis methods, the statistical accuracy of
ground-state quantities is also enhanced because correlation functions
computed at shorter Euclidean times can be used where the signal-to-noise is
greater.  As current lattice simulations are performed with ever-greater
resolution at short Euclidean times as lattice spacings are decreased toward
the continuum limit, simultaneous extraction of ground and excited-state
quantities will be essential to extract full value from such large-scale
(and expensive) computations.

A variety of analysis techniques have been applied to extracting the
excited-state spectrum from correlation functions. The most widely used method
is a nonlinear least-squares (NLLS) fit to a model function, such as a sum
over two or more exponentials.  Operationally, even such a simple nonlinear
fit can be fraught with difficulty, from establishing the range of Euclidean
times included in the dataset to stabilizing the convergence of minimization
algorithms by careful choices of initial guesses or temporarily freezing
selected fit parameters during the minimization process, all of which require
intervention by a trained expert.

At such times, the expert typically turns to black-box methods for guidance
because they do not require intervention to determine initial guesses and
fitting ranges:  estimates of correlation functions go into the black box and
estimates of hadron energies come out.  The main detraction of black-box
methods are that the produced estimates are expected to have larger
uncertainties, making them \textit{sub-optimal} relative to least squares
methods~\cite{Gauss:1823}.  Marrying the two approaches can effectively
combine the best features of both, leading to a highly-automated analysis
program producing optimal estimates of energies.  Prior to this work, black-box
methods were mainly useful for extracting ground-state and, perhaps, first
excited-state energies~\cite{Fleming:2004hs,Guadagnoli:2004wm}.  We believe
the method described below will provide the needed black-box method for
estimating as many energies from fixed set of correlations functions as are
likely to be extracted from a NLLS fit.

The structure of this paper is as follows: Sec.~\ref{sec:theory} provides
theoretical formulations of the excited-state effective masses for single and
multiple correlators; it also explores certain extensions to these techniques:
linear prediction and periodic boundary conditions. In Sec.~\ref{sec:data}, we
apply the methodology to some characteristic lattice correlators and compare
the results with simple fitting and the variational
method~\cite{Michael:1985ne,Luscher:1990ck}. Conclusions and future outlook
are given in Sec.~\ref{sec:summary}.  Numerous details and examples are
included in the appendices.  Preliminary details of this work were presented
in Ref.~\cite{Fleming:2006zz}.

\section{\label{sec:theory} Theoretical Basis}

The hadron spectrum can be calculated in LQCD using two-point hadronic
correlation functions
\begin{equation}
C(t_0, t) = \left\langle 0 \left| O(t)\ O^\dagger(t_0)
\right| 0 \right\rangle,
\end{equation}
where the creation and annihilation operators $O^\dagger$ and $O$
transform irreducibly under the symmetries of the lattice space group
\cite{Basak:2005aq,Basak:2005ir,Moore:2005dw,Moore:2006ng}.  After taking
momentum (and spin for baryons) projection and inserting a complete set of
hadronic eigenstates of the Hamiltonian (ignoring the variety of boundary
condition choices possible), this becomes
\begin{equation}
  \label{eq:2pt-model}
  C(\vec{p}, t_n) = \sum_{m=1}^M A_m(\vec{p})
  \exp\left[-(t_0 + n a) E_m(\vec{p})\right]
\end{equation} \[
  n \ge 0, \quad A_m,\ E_m \in \mathbb{R}, \quad
  0 \le E_1 \le E_2 \le \cdots \le E_M .
\]
where $A_m$ contains not only the overlap factor between the eigenstate and
states created by the operators but also any kinetic factors that do not
depend on Euclidean time separation $t_n = n a = t-t_0$.

\subsection{\label{sub:effmass}Effective Masses}

In general, a two-point correlation function computed on $N=2M$ time-slices
$t_n$ will admit an exact algebraic solution having the form of
Eq.~(\ref{eq:2pt-model}) with $M$ energies $E_m$ and amplitudes $A_m$.  The
problem to solve is the nonlinear system of equations $\bvec{y} = \bvec{V}(x)
\ \bvec{a}$
\begin{equation}
  \label{eq:VandermondeQCD}
  \left[\begin{array}{c}
  y_1    \\
  y_2    \\
  y_3    \\
  y_4    \\
  \vdots \\
  y_{2M}
  \end{array}\right] = \left[\begin{array}{cccc}
  1          & 1          & \cdots & 1          \\
  x_1        & x_2        & \cdots & x_M        \\
  x_1^2      & x_2^2      & \cdots & x_M^2      \\
  x_1^3      & x_2^3      & \cdots & x_M^3      \\
  \vdots     & \vdots     & \ddots & \vdots     \\
  x_1^{2M-1} & x_2^{2M-1} & \cdots & x_M^{2M-1}
  \end{array}\right] \left[\begin{array}{c}
  a_1    \\
  \vdots \\
  a_M
  \end{array}\right]
\end{equation}
for $x_m = \exp\left[-a E_m\left(\vec{p}\right)\right]$ and $a_m =
A_m\left(\vec{p}\right) \exp\left[ -t_0 E_m\left(\vec{p}\right)\right]$ where
$y_n = C\left(\vec{p},t_n\right)$. $\bvec{V}(x)$ is known as $2M \times M$
rectangular Vandermonde matrix.

By inspection, it appears the problem is of polynomial degree $2M$ and thus by
the Abel-Ruffini theorem~\cite{Ruffini:1799,Abel:1826} should not admit a
general closed form solution in terms of radicals for $M>2$. The $M=1$
solution is simple to compute and is widely known in the lattice QCD
literature as the \textsl{effective mass} solution. Note already that the
simple effective mass problem is linear and has only one solution, suggesting
that the polynomial degree is actually of order $M$.

The $M=2$ solution was explicitly constructed by one of the
authors~\cite{Fleming:2004hs} and was independently constructed some time
later by others~\cite{Guadagnoli:2004wm}. It was noted~\cite{Fleming:2004hs}
that the problem, when reduced, required only the solution of a quadratic
equation and so it was conjectured that the general problem of size $M$ could
be reduced to a polynomial equation in one variable of degree $M$.

An efficient algorithm has been available for some time for solving square
Vandermonde systems~\cite{Bjorck:1970} by making them upper triangular.  This
approach works equally well for rectangular Vandermonde systems as in
Eq.~(\ref{eq:VandermondeQCD}). Furthermore, this approach reveals why the
solution for the energies $E_m$ can be found without solving for the
amplitudes $A_m$ and why the problem is of polynomial degree $M$.

As a first step toward extracting the energies $E$ from our data, we transform
the system so that $V(x)$ is in upper triangular form~\cite{Bjorck:1970} by
pre-multiplying by the lower $2M \times 2M$ bi-diagonal matrices:
\begin{equation}
\label{eq:BP_prefactor}
L_m(x) = \left[\begin{array}{ccccccc}
1 &        &   &     &        &        &    \\
0 & \ddots &   &     &        &        &    \\
  & \ddots & 1 &     &        &        &    \\
  &        & 0 & 1   &        &        &    \\
  &        &   & x_m & -1     &        &    \\
  &        &   &     & \ddots & \ddots &    \\
  &        &   &     &        & x_m    & -1 \\
\end{array}\right]
\end{equation}
where the first $-1$ on the diagonal appears in the $m+1$ row and column. In
the appendices, we demonstrate in detail how the general solutions for $M=2$, 3
and 4 work.


Finding a general approach for $M > 4$ would be a tough challenge.  Although
Abel's Impossibility Theorem proves there are no general solutions in radicals
for polynomials higher than quartic order, there are numerical methods for
finding the roots of polynomials of any order. The general form for the
polynomial follows from Eqs.~(\ref{eq:V1x4x2determinant}),
(\ref{eq:V1x6x3determinant}) and (\ref{eq:V1x8x4determinant}) in the appendices:
\begin{equation}
\label{eq:V1x2MxMdeterminant}
\left\vert \bvec{\mathcal{H}} \right\vert =
\left\vert\begin{BMAT}(b){cccc.c}{cccc}
y_1     & y_2     & \cdots & y_M     & 1      \\
y_2     & y_3     & \cdots & y_{M+1} & x_1    \\
\vdots  & \vdots  & \ddots & \vdots  & \vdots \\
y_{M+1} & y_{M+2} & \cdots & y_{2M}  & x_1^M
\end{BMAT}\right\vert = 0.
\end{equation}

Prony~\cite{Prony:1795} showed that problems in the form of
Eq.~(\ref{eq:2pt-model}) implied the following system of equations $\bvec{y} =
\bvec{H}(y) \ \bvec{p}$
\begin{equation}
\label{eq:linear_prediction_hankel_system}
\left[ \begin{array}{c}
  y_1 \\ y_2 \\ \vdots \\ y_M
\end{array} \right] = - \left[ \begin{array}{ccc}
  y_2    & \cdots & y_{M+1} \\
  y_3    & \cdots & y_{M+2}   \\
  \vdots & \ddots & \vdots  \\
  y_{M+1}  & \cdots & y_{2M}
\end{array} \right] \left[ \begin{array}{c}
  p_1 \\ p_2 \\ \vdots \\ p_M
\end{array} \right]
\end{equation}
where the $M \times M$ matrix $\bvec{H}(y)$ has the special structure of a
Hankel matrix and the components $p_m$ of $\bvec{p}$ are the coefficients of a
polynomial
\begin{equation}
\label{eq:linear_prediction_coeff}
P(x) = \prod_{m=1}^M (x - x_m)
= 1 + \sum_{m=1}^M p_m x^m .
\end{equation}
The Prony-Yule-Walker method (or just Prony's method, for
short)~\cite{Prony:1795,Yule:1927,Walker:1931} solves
Eq.~(\ref{eq:linear_prediction_hankel_system}) to find the coefficients and
then finds the $M$ roots of the polynomial in
Eq.~(\ref{eq:linear_prediction_coeff}). The amplitudes are determined by
substituting the roots into Eq.~(\ref{eq:VandermondeQCD}) and solving it. Note
again that using $2M$ timeslices of correlation function data to determine $M$
effective masses is a problem of polynomial order $M$.

The general conditions under which the solutions of the Hankel and Vandermonde
systems coincide is presented in Ref.~\cite{Vandevoorde:1996}. Here we provide
a simple demonstration that both solutions are the same under the assumption
that there are no complications like degeneracies in the energy spectrum of
Eq.~(\ref{eq:2pt-model}). Assuming $\bvec{H}(y)$ is invertible, solving
Eq.~(\ref{eq:linear_prediction_hankel_system}) gives
\begin{equation}
\bvec{p} = \bvec{H}^{-1} \bvec{y}, \qquad
P(x) = 1 + \bvec{p}^\top \bvec{x}
     = 1 + \left( \bvec{H}^{-1} \bvec{y} \right)^\top \bvec{x}
\end{equation}
for the polynomial of Eq.~(\ref{eq:linear_prediction_coeff}) and where
$\bvec{x}^\top = \left( x, x^2, \cdots, x^M \right)$. Recall that the inverse
can be written in terms of the adjoint, or matrix of cofactors, $\bvec{H}^{-1}
= \bvec{C} / \left| \bvec{H} \right|$, $C_{ij} = (-1)^{i+j} \left|
\bvec{H}(i;j) \right|$, where the notation $\bvec{H}(i;j)$ means removing row
$i$ and column $j$. For Prony's method, we can rescale $P(x) \to \left|
\bvec{H} \right| P(x)$ and still find the roots $x_m$ by solving
\begin{equation}
  \left| \bvec{H} \right|
  + \left( \bvec{C}\,\bvec{y} \right)^\top \bvec{x} = 0
\end{equation}
for $x$.

Returning to the Vandermonde method, the determinant of
Eq.~(\ref{eq:V1x2MxMdeterminant}) can be expanded in terms of its cofactors
$\mathcal{C}_{ij} = (-1)^{i+j} \left| \bvec{\mathcal{H}}(i;j) \right|$
\begin{equation}
\left| \bvec{\mathcal{H}} \right| = (-1)^M \left| \bvec{H} \right|
  + \sum_{i=1}^M \mathcal{C}_{i+1,M+1} x^i
  = (-1)^M \left| \bvec{H} \right|
  + \sum_{i=1}^{M} (-1)^{i+M} \left| \bvec{\mathcal{H}}(i+1;M+1) \right| x^i .
\end{equation}
As usual, each cofactor can be expanded in terms of further cofactors where
additional rows and columns are removed:
\begin{equation}
\left| \bvec{\mathcal{H}}(i+1;M+1) \right| = \sum_{j=1}^M
(-1)^{j+1} \left| \bvec{\mathcal{H}}(1,i+1;j,M+1) \right| y_j.
\end{equation}
By eliminating the first row and last row of $\bvec{\mathcal{H}}$ in
Eq.~(\ref{eq:V1x2MxMdeterminant}) we recover $\bvec{H} =
\bvec{\mathcal{H}}(1;M+1)$ and for the cofactors
\begin{equation}
(-1)^{j+1} \left| \bvec{\mathcal{H}}(1,i+1; j,M+1) \right|
= (-1)^j \left| \bvec{H}(i;j) \right|
\end{equation}
so the desired identity is recovered
\begin{equation}
\left| \bvec{\mathcal{H}} \right| = (-1)^M \left| \bvec{H} \right|
+ (-1)^M \sum_{i=1}^M \sum_{j=1}^M (-1)^{i+j}
\left| \bvec{H}(i;j) \right| y_j x^i .
\end{equation}
up to a possible overall minus sign for odd $M$, which is irrelevant for
finding roots.  There is a unique set of solutions to the Vandermonde
and Hankel systems (under the assumption of noise-free correlation
functions with non-degenerate energy levels), so other considerations
should determine which is the better method to construct the polynomial. 
It is our experience that computing coefficients from
Eq.~(\ref{eq:V1x2MxMdeterminant}) is preferred to solving
Eq.~(\ref{eq:linear_prediction_hankel_system}) as statistical noise in
the correlation functions can lead to nearly singular Hankel matrices
which are difficult to invert.

In an earlier work~\cite{Fleming:2004hs}, one of the authors showed that
Prony's method (also called \textsl{linear prediction}) could easily be
extended to use more than $2M$ timeslices of a correlation function
to extract only $M$ masses by constructing an over-constrained system
of equations analogous to Eq.~(\ref{eq:linear_prediction_hankel_system}).
It is not obvious how to construct and solve a similar over-constrained
system in the Vandermonde case.  Thus, Prony's method has a potential
advantage that more time samples of the correlation function can
be used to extract the same number of energy levels leading to reduced
statistical fluctuations.

\subsection{\label{sub:multicor}Solutions with multiple correlation functions}

When constructing correlation functions in LQCD, care is taken to ensure that
the correlation function transforms irreducibly under the symmetries of the
lattice space group~\cite{Basak:2005aq,Basak:2005ir,Moore:2005dw,
Moore:2006ng}. For the model function, as in Eq.~(\ref{eq:2pt-model}), this
implies that the amplitudes depend on the details of the specific correlation
function but that the energies depend only on the irreducible representation.
Since it is common in lattice QCD simulations to compute at least two distinct
correlation functions for each irreducible representation, effective mass
solutions which combine data from multiple correlations are also possible.

Assume that there are $K$ correlation functions available as in
Eq.~(\ref{eq:2pt-model}) that differ only in their amplitudes:
\begin{equation}
  \label{eq:multiple_hadron_correlations}
  C_k(\vec{p}, t_n) = \sum_{m=1}^M A_{km}(\vec{p})
  \exp\left[-(t_0 + n a) E_m(\vec{p})\right]
\end{equation} \[
  n \ge 0, \quad A_{km},\ E_m \in \mathbb{R}, \quad
  0 \le E_1 \le E_2 \le \cdots \le E_M .
\]
Data from the same $N$ time slices will be used in the following from each
correlation function to construct $M$ effective masses. Under this assumption
the condition that there will be equal number of data points as unknowns is $K
N = (K + 1) M$. In Appendix~\ref{sec:Meff_m234} are the three solutions that
satisfy the condition for $K=1$, up to quartic order. There are four more
solutions (up to quartic order) for $K > 1$: $(K,M,N) = (2,2,3)$, $(2,4,6)$,
$(3,3,4)$ and $(4,4,5)$ (demonstrated in Appendix~\ref{sec:KMN-examples}).
Relaxing the assumption that the same number of time slices are used from each
correlation function will allow for more possibilities up to quartic order. It
is straightforward to generalize to these cases if desired.

The general form of the polynomial equation can be inferred by studying the
solved examples in Eqs.~(\ref{eq:V2x2x3determinant}),
(\ref{eq:V2x4x6determinant}), (\ref{eq:V3x3x4determinant}) and
(\ref{eq:V4x4x5determinant}). Define $K$ Hankel matrices $H_k^{N\times M_k}$
for each of the correlation functions with the constraints $\sum_{k=1}^K M_k =
M$ and $N=M+1$. Then the general form of the polynomial equation is
\begin{equation}
  \label{eq:VKxMxNdeterminant}
  \left| \begin{BMAT}(b){c.c.c.c.c}{c}
    H_1^{N \times M_1} &
    H_2^{N \times M_2} &
    \cdots             &
    H_M^{N \times M_K} &
    \begin{array}{c}
      1 \\ x_1 \\ x_1^2 \\ \vdots \\ x_1^M
    \end{array}
  \end{BMAT}\right| = 0.
\end{equation}
As previously discussed, each Hankel matrix is generally of full column rank
and, if the correlation functions are linearly independent, then the columns
of different Hankel matrices are also linearly independent. So,
Eq.~(\ref{eq:VKxMxNdeterminant}) will only be satisfied for discrete values of
$x_1$ corresponding to the roots of the polynomial.

\subsection{\label{sub::PBC}Periodic Boundary Conditions}

In practical LQCD calculations, the temporal extent is finite so the choice of
temporal boundary conditions affects hadronic correlation functions near the
boundary. For simplicity, starting from Eq.~(\ref{eq:2pt-model}), set $t_0 =
0$ and identify the points $t_0 = 0$ and $t_N = N a$ which can be done using
modular arithmetic, \textit{i.e.}\ $t_n = (n \bmod N) a$. For anti-periodic
boundary conditions, the typical hadronic Euclidean time correlation function
is described by the model function
\begin{equation}
  \label{eq:ap_hadron_correlation}
  C(\vec{p}, t_n) = \sum_{m=1}^M \left\{
    A_m(\vec{p}) \exp\left[-(n \bmod N) a E_m(\vec{p})\right]
    + (-1)^B A_m^*(\vec{p}) \exp\left[-(N - n \bmod N) a E_m^*(\vec{p})\right]
  \right\}
\end{equation} \[
  n \ge 0, \quad A_m,\ A_m^*,\ E_m,\ E_m^* \in \mathbb{R}, \quad
  0 \le E_1   \le E_2   \le \cdots \le E_M   , \quad
  0 \le E_1^* \le E_2^* \le \cdots \le E_M^*.
\]
For periodic boundary conditions, set $(-1)^B \to 1$. For mesons, $B=0$ but
more importantly $A_m = A_m^*$ and $E_m = E_m^*$, which is not true for baryons
($B=1$). So, baryon correlation functions represent $M$ states propagating to
the left and $M$ different states propagating to the right for a total of $2M$
states.

Meson correlation functions represent the same $M$ states propagating to the
right and left.  However, time-reversal symmetry requires $C(\vec{P},t_n) =
C(\vec{P},t_{N-n})$, up to noise terms, so that only half of the computed
timeslices are truly independent. Thus, as was the case with baryons,
information about only $M$ states in any given quantum number channel can be
extracted from a single correlation function computed on $2M$ timeslices in a
finite box.  As shown in Ref.~\cite{Fleming:2004hs}, this can be made explicit
by writing the meson correlation function as
\begin{equation}
C(\tau_n) = \sum_{m=1}^M A_m \exp(-a N E_m / 2 ) \cosh( a n E_m ), \quad
\tau_n = ( n - N/2) a.
\end{equation}
To write this result in the Vandermonde form of Eq.~(\ref{eq:VandermondeQCD}),
define the variables
\begin{equation}
a_m = A_m \exp(-a N E_m/2), \quad x_m = \cosh(a E_m), \quad y_n =
\frac{1}{2^{n-1}} \sum_{j=0}^{n-1} \left(
  \begin{array}{c} n-1 \\ j \end{array}
\right) C(\tau_{n-2j-1}).
\end{equation}

When solving Eq.~(\ref{eq:VandermondeQCD}), the domain of the solutions $x_m$
will be the real numbers or complex conjugate pairs since real-valued
correlation functions are used as input.  Complex-valued solutions are clearly
unphysical and should be discarded as they are likely due to noise.  Real
solutions may also be unphysical if they cannot be used to extract a
non-negative energy, and this will depend on the details of the model function
and hence the boundary conditions.  For example, for the basic model of
Eq.~(\ref{eq:2pt-model}), only the solutions $0 < x_m \le 1$ will yield
non-negative energies.  For mesons in periodic boxes, $x_m = \cosh(a E_m)$ so
only $x_m \ge 1$ will yield non-negative energies.  Finally, for baryons in
periodic boxes, the $M$ states propagating to the right have $x_m = \exp(-a
E_m)$ and the $M$ states propagating to the left have $x_{M+m} = \exp(a
E_m^*)$ so all solutions $x_m > 0$ are physical and $x_m > 1$ means the state
is propagating to the left.

For some lattice fermion actions, \textit{e.g.}\ staggered
\cite{Sharatchandra:1981si,vandenDoel:1983mf,Creutz:1986ky} or domain-wall
fermions~\cite{Syritsyn:2007mp,WalkerLoud:2008bp}, a variation of
Eq.~(\ref{eq:2pt-model}) is needed as a starting point to account for states
which oscillate in time.  For example, an appropriate model for staggered
mesons on an infinite lattice is
\begin{equation}
  \label{eq:stag_meson_correlation}
  C(\vec{p}, t_n) = \sum_{m=1}^M \left\{
    A_m(\vec{p}) \exp\left[-n a E_m(\vec{p})\right]
    + (-1)^n A_m^*(\vec{p}) \exp\left[-n a E_m^*(\vec{p})\right]
  \right\}
\end{equation} \[
  n \ge 0, \quad A_m,\ A_m^*,\ E_m,\ E_m^* \in \mathbb{R}, \quad
  0 \le E_1   \le E_2   \le \cdots \le E_M   , \quad
  0 \le E_1^* \le E_2^* \le \cdots \le E_M^*.
\]
There are two independently ordered sets of states, half of which oscillate as
$(-1)^n$.  When solving Eq.~(\ref{eq:VandermondeQCD}), such oscillating
solutions will have physical solutions if $-1 \le x_m < 0$.  Similarly, for
staggered baryons with periodic boundary conditions, physical solutions with
$x_m \le -1$ are certainly expected as oscillating states moving to the left.

Generally speaking, model functions appropriate for the common lattice
discretizations and choice of boundary conditions can be formulated and
rewritten in the Vandermonde form of Eq.~(\ref{eq:VandermondeQCD}).  The
physical interpretation of the solutions $x_m$ depends on the details of the
discretization and boundary conditions.  In some cases, all real solutions may
have a physical interpretation and thus cannot be immediately discarded
without further statistical analysis.

\section{\label{sec:data}Numerical Results}

Although the ground-state effective mass has a long tradition of use in the
lattice QCD community, not much work has examined excited states and fewer yet
their effective masses. In this section, we will demonstrate the application
of these effective-mass techniques to some typical lattice correlation
functions. At the end, we compare the results with the ones from variational
method.

\begin{figure}
\includegraphics[width=0.7\textwidth]{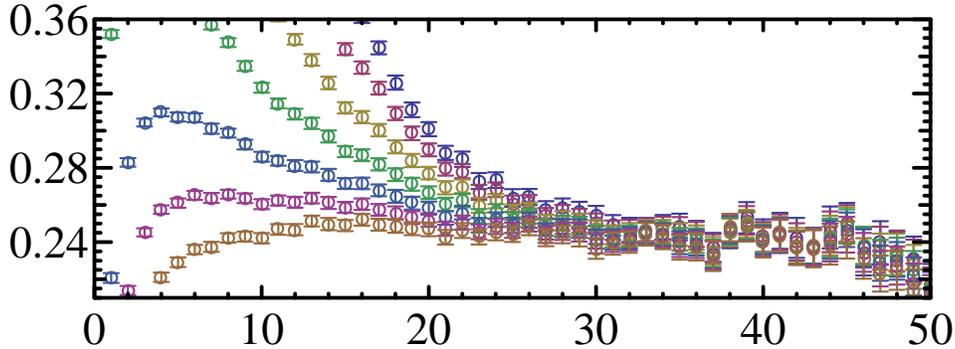}
\caption{Effective mass plots from the 7 smeared-point proton correlators used
in this work. The horizontal axis shows time and the vertical axis shows the
effective masses, both in lattice units.}
\label{fig:meff}
\end{figure}

The data on which we demonstrate these methods is from a study done using the
quenched approximation to QCD, \textit{i.e.}\ where the effects of quantum
fluctuations of quark-antiquark pairs in the vacuum are ignored, greatly
reducing the computational cost but leading to an unknown, but hopefully
small, systematic error.  We generated an ensemble of $16^3\times64$
anisotropic lattices with Wilson gauge action and nonperturbative clover
fermion action using Dirichlet boundary condition. The spatial lattice spacing
$a_s$ is about 0.125~fm with anisotropy 3 (that is, temporal spacing $a_t^{-1}
\approx 6$~GeV). The parameters used in the fermion action give a
720-MeV~pion. Specifically, our data are proton correlators using 7 Gaussian
smearing parameters, 0.5--6.5 in steps of 1.0, including both smeared-point
and smeared-smeared source-sink combinations.  Fig.~\ref{fig:meff} shows the
single-state effective-mass plots for these smeared-point proton correlators. 

\subsection{\label{sub:effmass-data}Excited-Effective Masses}

\begin{figure}
\includegraphics[width=\textwidth]{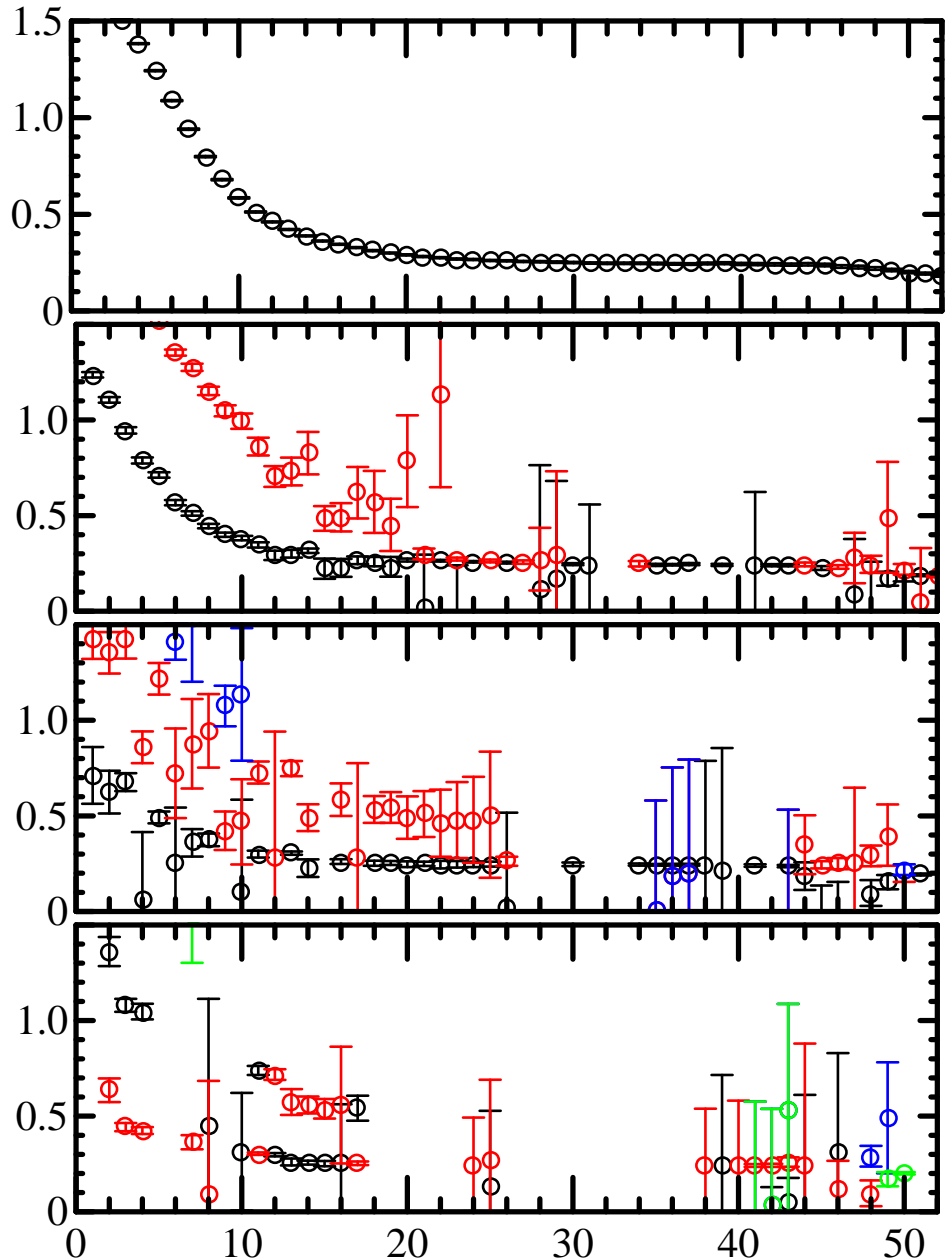}
\caption{Higher-effective mass plots with (top-to-bottom) one, two, three and
four masses. The colors indicate the (black) ground, (red) first-excited,
(blue) second-excited and (green) third-excited states.}
\label{fig:multieff}
\end{figure}

We now apply the excited-effective masses to our nucleon data. In
Fig.~\ref{fig:multieff} we show the results of applying the single-correlator
($K=1$) excited-effective masses to the nucleon data for all $M$ with analytic
solutions in terms of radicals. Notice that as the number of states included
increases, the amount of early-time contamination is decreased. As the
formulae better account for the exact form of the correlator, more of the time
range can be reasonably used to determine the states. However, as the number
of roots increases, the occurrence of ``bad'' roots (those that are negative
or imaginary) tends to increase as well. Since these have to be thrown out,
this causes gaps in the extracted states where the results are unreliable.

\begin{figure}
\includegraphics[width=\textwidth]{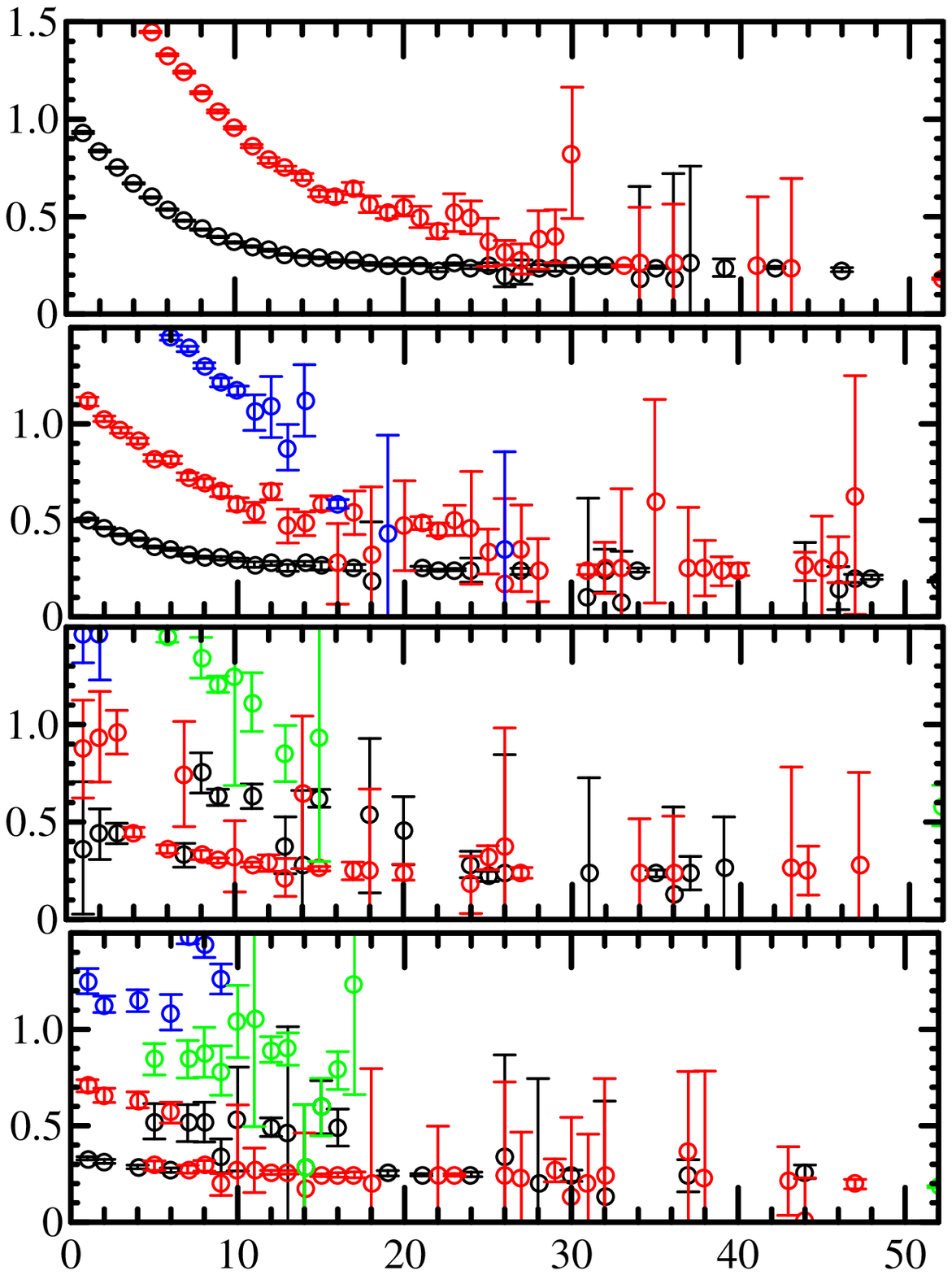}
\caption{Multiple-correlator higher-effective mass plots with (top-to-bottom)
$\{K,M\}$ equal to $\{2,2\}$, $\{3,3\}$, $\{4,2\}$ and $\{4,4\}$. The colors
indicate the (black) ground, (red) first-excited, (blue) second-excited and
(green) third-excited states.}
\label{fig:multicor}
\end{figure}

In Fig.~\ref{fig:multicor} we show the results of applying the multiple
correlator excited-effective masses to the nucleon data for all combinations
of $K$ and $M$ that have solutions in terms of radicals. Notice that as the
number of correlators increases, the quality of the extracted masses improves.
The gaps in the data where bad roots appear become much less noticeable.

\begin{figure}
\includegraphics[width=0.7\textwidth]{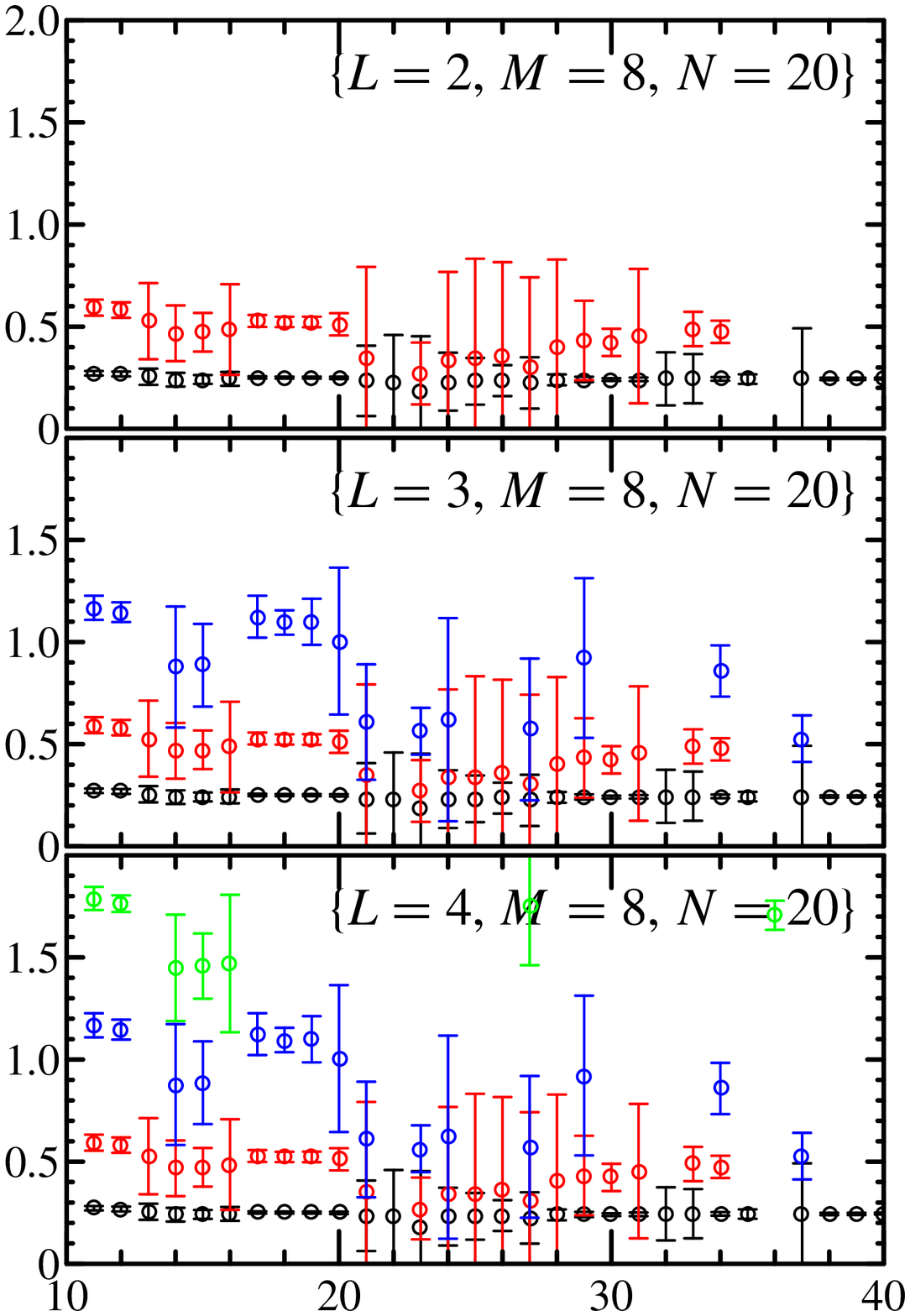}
\caption{Effective-mass plots from the linear prediction black-box method at
fixed parameters $N=20$ and $M=8$}
\label{fig:LPP-meff}
\end{figure}

We demonstrate this approach using the smallest smearing parameter (0.5)
smeared-point correlator. There are a few parameters in the linear prediction
method which we can tune: the number of desired states $L$, the number of time
slices used to predict the later time point $N$, and the order of the
polynomial $M$. In this work, we will show a selection of the better choices
in these degrees of freedom. Figure~\ref{fig:LPP-meff} shows the effective
mass plot for $L=2,3,4$ from a single Gaussian smeared-point correlator with
fixed parameters $N=20$ and $M=8$. The excited states are consistent with each
other as one increases the value of $K$. Since we have used a large value of
$N$ to form the polynomial, each point uses information extracted from 20
timeslices. Thus, one does not need a large plateau to determine the final
mass. One also notes that since we only use a single correlator to extract
multiple states, the multiple states will be correlated; that is, large errors
on higher-excited states will make the ground state noisy as well. A future
improvement would naturally be to extend this approach to multiple
correlators.

\subsection{\label{sub:variational}Variational Method}

Variational method~\cite{Michael:1985ne,Luscher:1990ck} is a powerful
tool for extracting multi-excited states in lattice QCD.  We construct an $r
\times r$ spectrum correlation matrix, $C_{ij}(t)$, where each element of the
matrix is a correlator composed from different smeared sources or operators
${\cal O}_i$ and ${\cal O}_j$. Then we consider the generalized eigenvalue
problem
\begin{eqnarray} 
C(t)\psi=\lambda(t,t_0)C(t_0)\psi, 
\end{eqnarray}
where $t_0$ determines the range of validity of our extraction of the lowest
$r$ eigenstates. If $t_0$ is too large, the highest-lying states will have
exponentially decreased too far to have good signal-to-noise ratio; if $t_0$
is too small, many states above the $r$ we can determine will contaminate our
extraction. Over some intermediate range in $t_0$, we should find consistent
results.

If the eigenvector for this system is $|\alpha\rangle$, and $\alpha$ goes from
1 to $r$. Thus the correlation matrix can be approximated as
\begin{eqnarray}
C_{ij}=\sum_{n=1}^r v_i^{n*} v_j^n e^{-tE_n}
\end{eqnarray}
with eigenvalues
\begin{eqnarray}
\lambda_n(t,t_0)=e^{-(t-t_0)E_n}
\end{eqnarray}
by solving
\begin{eqnarray}
C(t_0)^{-1/2}C(t)C(t_0)^{-1/2}\psi=\lambda(t,t_0)\psi.
\end{eqnarray}
Further analysis on the principal correlators, $\lambda_n(t,t_0)$, reveals
information on the energy levels, $E_n$.

\begin{figure}
\includegraphics[width=0.7\textwidth]{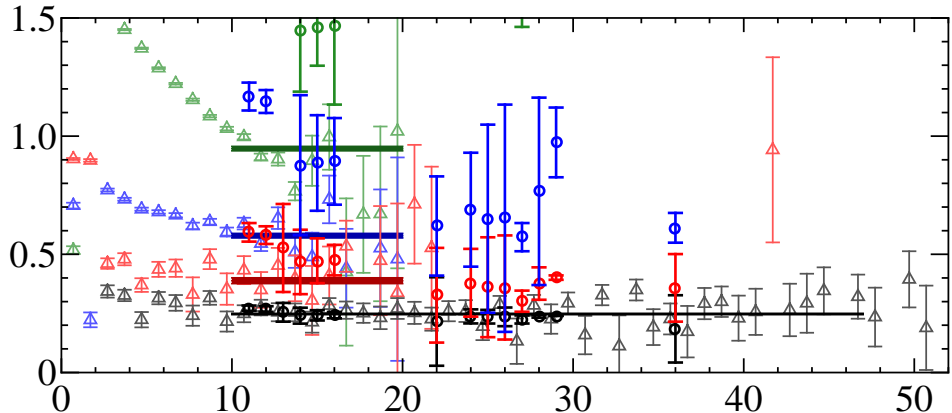}
\caption{Comparison between the variational method ($4 \times 4$) and linear
prediction with 4 extracted states}
\label{fig:comp-meff}
\end{figure}

The results from the linear prediction approach in Sec.~\ref{sub:effmass} are
compared with the variational method ($4 \times 4$ with smearing parameter
ranging 0.5--3.5), as shown in Fig.~\ref{fig:comp-meff}. Here we shift time
with respect to the linear prediction plot by 10 to have a better comparison
with the plateau region from the variational method. For the ground state, the
numbers are consistent with the result from the variational approach,
including the size of the error bar.  This is remarkable, given that the
amount of input information is a factor of 16 less in the linear prediction
approach. The first-excited state is consistent but has larger error bar,
which is no surprise. As for the second-excited state, it seems to be
consistent with the variational ones but definitely needs more statistics. The
third-excited state is much larger than expected from the variational
approach, which might be caused by contamination from even higher excited
states.

\section{\label{sec:summary}Conclusions}
The determination of the physical properties of the excited-state hadrons is
currently of great interest due to the construction of the 12 GeV upgrade at
Jefferson Lab, where such properties will be measured experimentally.  Lattice
QCD methods have the potential to predict these properties, provided they can
be extracted from the exponential time series, called correlation functions,
computed in Monte Carlo simulations.  In this work, we demonstrate a powerful,
yet easy to use, black-box method for analyzing one or more correlation
functions and extracting information about excited states.  It can easily
be adapted to various choices of boundary conditions and discretizations.
While the method can be used by itself to estimate physical properties
of excited hadrons, we anticipate that it will also be useful as a method
for generating initial guesses for nonlinear least-squares minimizers.

\section*{Acknowledgments}

Computations were performed on clusters at Jefferson Lab with time awarded
under the U.S.\ DOE's National Computational Infrastructure for Lattice Gauge
Theory (USQCD), using the Chroma software suite~\cite{Edwards:2004sx}
developed under DOE's Scientific Discovery through Advanced Computing
initiative. This work was supported by U.S.\ NSF grants PHY-0556243 and
PHY-0801068.  Authored by Jefferson Science Associates, LLC under U.S. DOE
Contract No. DE-AC05-06OR23177. The U.S. Government retains a non-exclusive,
paid-up, irrevocable, world-wide license to publish or reproduce this
manuscript for U.S. Government purposes. 

\appendix

\section{\label{sec:Meff_m234} General effective mass solution for
         $M=2$, 3, and 4}

\subsection{\label{sub:M=2} General solution for $M=2$}

To reduce the $M=2$ problem, we pre-multiply by two factors of the bi-diagonal
matrices of Eq.~(\ref{eq:BP_prefactor}) to find the reduced equation $L_2 L_1
y = L_2 L_1 V a$. By introducing the auxiliary quantities:
\begin{eqnarray}
\label{eq:alpha}
\alpha_i & = & x_1 y_{i-1}      - y_i      \quad ( 2 \le i \le 2M ) \\
\label{eq:beta}
\beta_j  & = & x_2 \alpha_{j-1} - \alpha_j \quad ( 3 \le j \le 2M )
\end{eqnarray}
the reduced system becomes:
\begin{eqnarray}
\label{eq:V1x4x2eq1}
y_1      & = & a_1 + a_2                    \\
\label{eq:V1x4x2eq2}
\alpha_2 & = & a_2 \left( x_1 - x_2 \right) \\
\label{eq:V1x4x2eq3}
\beta_3  & = & 0                            \\
\label{eq:V1x4x2eq4}
\beta_4  & = & 0
\end{eqnarray}
The first half of the equations, Eqs.~(\ref{eq:V1x4x2eq1}) and
(\ref{eq:V1x4x2eq2}), involve both the amplitudes $A_1, A_2$ and the energies
$E_1, E_2$ but the second half involve only the energies. It will be true for
any $M$, in general, that the last $M$ equations can be solved first to find
all the energies. Once all the energies are known, the first $M$ equations
form a square upper triangular system that can be solved efficiently by
backward substitution to find the amplitudes.

To see that Eqs.~(\ref{eq:V1x4x2eq1})--(\ref{eq:V1x4x2eq4}) yield the known
solution \cite{Fleming:2004hs}, first substitute Eq.~(\ref{eq:beta}) and
eliminate $x_2$ from Eqs.~(\ref{eq:V1x4x2eq3})--(\ref{eq:V1x4x2eq4}) to find
\begin{equation}
\label{eq:V1x4x2reduction1}
\alpha_2 \alpha_4 - \alpha_3^2 = 0, \quad \mathrm{or} \quad
\left\vert \begin{array}{cc}
  \alpha_2 & \alpha_3 \\
  \alpha_3 & \alpha_4
\end{array} \right\vert = 0
\end{equation}
where we note that the l.h.s.\ is the determinant of a $2 \times 2$ Hankel
matrix or perhaps the minor of a larger Hankel matrix. After substituting
Eq.~(\ref{eq:alpha}), this gives the known quadratic equation
\begin{equation}
\label{eq:V1x4x2quadratic}
\left( y_2^2 - y_1 y_3   \right) x_1^2 +
\left( y_1 y_4 - y_2 y_3 \right) x_1   +
\left( y_3^2 - y_2 y_4   \right)       = 0 .
\end{equation}
Note that this can also be written
\begin{equation}
\left\vert
  \begin{array}{cc} y_1 & y_2 \\ y_2 & y_3 \end{array}
\right\vert x_1^2
- \left\vert
  \begin{array}{cc} y_1 & y_2 \\ y_3 & y_4 \end{array}
\right\vert x_1
+ \left\vert
  \begin{array}{cc} y_2 & y_3 \\ y_3 & y_4 \end{array}
\right\vert = 0
\end{equation}
where the coefficients are not determinants of Hankel matrices but minors of a
single Hankel matrix. So, it can be written even more compactly as
\begin{equation}
\left\vert\begin{BMAT}(b){cc.c}{ccc}
\label{eq:V1x4x2determinant}
y_1 & y_2 & 1     \\
y_2 & y_3 & x_1   \\
y_3 & y_4 & x_1^2
\end{BMAT}\right\vert = 0
\end{equation}
where the left block is a Hankel matrix and the right block is a Vandermonde
matrix.

\subsection{\label{sub:M=3} General solution for $M=3$}

Using the auxiliary quantities defined in Eqs.~(\ref{eq:alpha}) and
(\ref{eq:beta}) and a third auxiliary quantity:
\begin{equation}
\label{eq:gamma}
\gamma_i = x_3 \beta_{i-1} - \beta_i \quad ( 4 \le i \le 2M )
\end{equation}
the reduced system of equations for $M=3$ is:
\begin{eqnarray}
\label{eq:V1x6x3eq1}
y_1      & = & a_1 + a_2 + a_3                                             \\
\label{eq:V1x6x3eq2}
\alpha_2 & = & a_2 \left( x_1 - x_2 \right) + a_3 \left( x_1 - x_3 \right) \\
\label{eq:V1x6x3eq3}
\beta_3  & = & a_3 \left( x_2 - x_3 \right) \left( x_1 - x_3 \right)       \\
\label{eq:V1x6x3eq4}
\gamma_4 & = & 0                                                           \\
\label{eq:V1x6x3eq5}
\gamma_5 & = & 0                                                           \\
\label{eq:V1x6x3eq6}
\gamma_6 & = & 0
\end{eqnarray}
Following the procedure of Sec.~\ref{sub:M=2}, substitute Eq.~(\ref{eq:gamma})
into the last three equations to eliminate $x_3$ and find the (redundant) set
of equations
\begin{equation}
\label{eq:V1x6x3reduction1}
\beta_3 \beta_5 - \beta_4^2       = 0 , \quad
\beta_3 \beta_6 - \beta_4 \beta_5 = 0 , \quad
\beta_4 \beta_5 - \beta_5^2       = 0
\end{equation}
or equivalently
\begin{equation}
\left\vert \begin{array}{cc}
  \beta_3 & \beta_4 \\
  \beta_4 & \beta_5
\end{array} \right\vert = 0 , \quad
\left\vert \begin{array}{cc}
  \beta_3 & \beta_4 \\
  \beta_5 & \beta_6
\end{array} \right\vert = 0 , \quad
\left\vert \begin{array}{cc}
  \beta_4 & \beta_5 \\
  \beta_5 & \beta_6
\end{array} \right\vert = 0 .
\end{equation}
Note the l.h.s.\ of these three equations are the same as the three
coefficients of Eq.~(\ref{eq:V1x4x2quadratic}) after the substitution $y_i \to
\beta_{i+2}$ and thus are minors of a Hankel matrix. Next, substitute
Eq.~(\ref{eq:beta}) into these equations to eliminate $x_2$ and find the
equation
\begin{equation}
\label{eq:V1x6x3reduction2}
\alpha_2 \alpha_4 \alpha_6 + 2 \alpha_3 \alpha_4 \alpha_5
- \alpha_4^3 - \alpha_3^2 \alpha_6 - \alpha_2 \alpha_5^2 = 0
\quad \mathrm{or} \quad
\left\vert \begin{array}{ccc}
  \alpha_2 & \alpha_3 & \alpha_4 \\
  \alpha_3 & \alpha_4 & \alpha_5 \\
  \alpha_4 & \alpha_5 & \alpha_6
\end{array} \right\vert = 0 .
\end{equation}
Finally, substituting Eq.~(\ref{eq:alpha}) will produce a cubic equation in
$x_1$:
\begin{eqnarray}
\label{eq:V1x6x3cubic}
\lefteqn{A x_1^3 + B x_1^2 + C x_1 + D = 0} \\
& & A = y_3^3 - 2 y_2 y_3 y_4 + y_1 y_4^2 + y_2^2 y_5 - y_1 y_3 y_5
  \nonumber \\
& & B = - y_3^2 y_4 + y_2 y_4^2 + y_2 y_3 y_5 - y_1 y_4 y_5 - y_2^2 y_6
        + y_1 y_3 y_6 \nonumber \\
& & C = y_3 y_4^2 - y_3^2 y_5 - y_2 y_4 y_5 + y_1 y_5^2 + y_2 y_3 y_6
        - y_1 y_4 y_6 \nonumber \\
& & D = - y_4^3 + 2 y_3 y_4 y_5 - y_2 y_5^2 - y_3^2 y_6 + y_2 y_4 y_6 .
  \nonumber
\end{eqnarray}
The coefficients $A$ through $D$ are minors of a Hankel matrix and so
Eq.~(\ref{eq:V1x6x3cubic}) can also be written compactly as
\begin{equation}
\label{eq:V1x6x3determinant}
\left\vert\begin{BMAT}(b){ccc.c}{cccc}
y_1 & y_2 & y_3 & 1     \\
y_2 & y_3 & y_4 & x_1   \\
y_3 & y_4 & y_5 & x_1^2 \\
y_4 & y_5 & y_6 & x_1^3
\end{BMAT}\right\vert = 0.
\end{equation}
The cubic equation can be solved using the method of Scipione del Ferro and
Tartaglia \cite{Cardano:1545}.

\subsection{\label{sub:M=4} General solution for $M=4$}

Defining a fourth auxiliary quantity:
\begin{equation}
\label{eq:delta}
\delta_i = x_4 \gamma_{i-1} - \gamma_i \quad ( 5 \le i \le 2M )
\end{equation}
the reduced system of equations for $M=4$ is:
\begin{eqnarray}
\label{eq:V8x4eq1}
y_1      & = & a_1 + a_2 + a_3 + a_4                                       \\
\label{eq:V8x4eq2}
\alpha_2 & = & a_2 \left( x_1 - x_2 \right) + a_3 \left( x_1 - x_3 \right)
               + a_4 \left( x_1 - x_4 \right)                              \\
\label{eq:V8x4eq3}
\beta_3  & = &   a_3 \left( x_1 - x_3 \right) \left( x_2 - x_3 \right)
               + a_4 \left( x_1 - x_4 \right) \left( x_2 - x_4 \right)     \\
\label{eq:V8x4eq4}
\gamma_4 & = & a_4 \left( x_1 - x_4 \right) \left( x_2 - x_4 \right)
               \left( x_3 - x_4 \right)                                    \\
\label{eq:V8x4eq5}
\delta_5 & = & 0                                                           \\
\label{eq:V8x4eq6}
\delta_6 & = & 0                                                           \\
\label{eq:V8x4eq7}
\delta_7 & = & 0                                                           \\
\label{eq:V8x4eq8}
\delta_8 & = & 0
\end{eqnarray}
Following the now familiar procedure, substitute Eq.~(\ref{eq:delta}) into the
last four equations to eliminate $x_4$ and find the (redundant) set of
equations:
\begin{eqnarray}
\gamma_5^2 - \gamma_4 \gamma_6 = 0        , &
\gamma_4 \gamma_7 - \gamma_5 \gamma_6 = 0 , &
\gamma_6^2 - \gamma_5 \gamma_7 = 0        , \\
\gamma_6^2 - \gamma_4 \gamma_8 = 0        , &
\gamma_5 \gamma_8 - \gamma_6 \gamma_7 = 0 , &
\gamma_7^2 - \gamma_6 \gamma_8 = 0          \nonumber
\end{eqnarray}
or as minors of a Hankel matrix:
\begin{eqnarray}
\left\vert \begin{array}{cc}
  \gamma_4 & \gamma_5 \\
  \gamma_5 & \gamma_6
\end{array} \right\vert = 0  , &
\left\vert \begin{array}{cc}
  \gamma_4 & \gamma_5 \\
  \gamma_6 & \gamma_7
\end{array} \right\vert = 0  , &
\left\vert \begin{array}{cc}
  \gamma_5 & \gamma_6 \\
  \gamma_6 & \gamma_7
\end{array} \right\vert = 0  , \\
\left\vert \begin{array}{cc}
  \gamma_4 & \gamma_6 \\
  \gamma_6 & \gamma_8
\end{array} \right\vert = 0  , &
\left\vert \begin{array}{cc}
  \gamma_5 & \gamma_6 \\
  \gamma_7 & \gamma_8
\end{array} \right\vert = 0  , &
\left\vert \begin{array}{cc}
  \gamma_6 & \gamma_7 \\
  \gamma_7 & \gamma_8
\end{array} \right\vert = 0  . \nonumber
\end{eqnarray}
Substitute Eq.~(\ref{eq:gamma}) and eliminate $x_3$ to find the next set of
(redundant) set of equations:
\begin{eqnarray}
\beta_5^3 - 2 \beta_4 \beta_5 \beta_6 + \beta_3 \beta_6^2 + \beta_4^2 \beta_7
- \beta_3 \beta_5 \beta_7 & = & 0 \\
- \beta_5^2 \beta_6 + \beta_4 \beta_6^2 + \beta_4 \beta_5 \beta_7
- \beta_3 \beta_6 \beta_7 - \beta_4^2 \beta_8 + \beta_3 \beta_5 \beta_8
& = & 0 \nonumber \\
\beta_5 \beta_6^2 - \beta_5^2 \beta_7 - \beta_4 \beta_6 \beta_7
+ \beta_3 \beta_7^2 + \beta_4 \beta_5 \beta_8 - \beta_3 \beta_6 \beta_8
& = & 0 \nonumber \\
- \beta_6^3 + 2 \beta_5 \beta_6 \beta_7 - \beta_4 \beta_7^2
- \beta_5^2 \beta_8 + \beta_4 \beta_6 \beta_8 & = & 0 \nonumber
\end{eqnarray}
or as minors of a Hankel matrix:
\begin{eqnarray}
\left\vert \begin{array}{ccc}
  \beta_3 & \beta_4 & \beta_5 \\
  \beta_4 & \beta_5 & \beta_6 \\
  \beta_5 & \beta_6 & \beta_7 \\
\end{array} \right\vert = 0, \quad
\left\vert \begin{array}{ccc}
  \beta_3 & \beta_4 & \beta_5 \\
  \beta_4 & \beta_5 & \beta_6 \\
  \beta_6 & \beta_7 & \beta_8 \\
\end{array} \right\vert = 0, \quad
\left\vert \begin{array}{ccc}
  \beta_3 & \beta_4 & \beta_5 \\
  \beta_5 & \beta_6 & \beta_7 \\
  \beta_6 & \beta_7 & \beta_8 \\
\end{array} \right\vert = 0, \quad
\left\vert \begin{array}{ccc}
  \beta_4 & \beta_5 & \beta_6 \\
  \beta_5 & \beta_6 & \beta_7 \\
  \beta_6 & \beta_7 & \beta_8 \\
\end{array} \right\vert = 0.
\end{eqnarray}
Again, note that the LHS of these four equations are the same as the four
coefficients of Eq.~(\ref{eq:V1x6x3cubic}) after the same substitution $y_i
\to \beta_{i+2}$. Next, substitute Eq.~(\ref{eq:beta}) and eliminate $x_2$ to
find the equation
\begin{eqnarray}
\lefteqn{
    \alpha_5^4
- 3 \alpha_4   \alpha_6   \alpha_5^2
- 2 \alpha_3   \alpha_7   \alpha_5^2
-   \alpha_2   \alpha_8   \alpha_5^2
+ 2 \alpha_3   \alpha_6^2 \alpha_5
+ 2 \alpha_4^2 \alpha_7   \alpha_5
} \\*
& &
+ 2 \alpha_2   \alpha_6   \alpha_7   \alpha_5
+ 2 \alpha_3   \alpha_4   \alpha_8   \alpha_5
-   \alpha_2   \alpha_6^3
+   \alpha_4^2 \alpha_6^2
+   \alpha_3^2 \alpha_7^2
\nonumber \\*
& &
-   \alpha_2   \alpha_4   \alpha_7^2
- 2 \alpha_3   \alpha_4   \alpha_6   \alpha_7
-   \alpha_4^3 \alpha_8
-   \alpha_3^2 \alpha_6   \alpha_8
+   \alpha_2   \alpha_4   \alpha_6   \alpha_8
= 0 \nonumber
\end{eqnarray}
As in Eqs.~(\ref{eq:V1x4x2reduction1}) and (\ref{eq:V1x6x3reduction2}) the
l.h.s.\ can be written as a determinant of a Hankel matrix of $\alpha_i$'s:
\begin{equation}
\left\vert \begin{array}{cccc}
  \alpha_2 & \alpha_3 & \alpha_4 & \alpha_5 \\
  \alpha_3 & \alpha_4 & \alpha_5 & \alpha_6 \\
  \alpha_4 & \alpha_5 & \alpha_6 & \alpha_7 \\
  \alpha_5 & \alpha_6 & \alpha_7 & \alpha_8
\end{array} \right\vert = 0 .
\end{equation}
Finally, substituting Eq.~(\ref{eq:alpha}) produces a quartic equation in
$x_1$:
\begin{eqnarray}
\label{eq:V1x8x4quartic}
\lefteqn{A x_1^4 + B x_1^3 + C x_1^2 + D x_1 + E = 0} \\
& A=\!\!\!\! & y_4^4 - 3 y_3 y_5 y_4^2 - 2 y_2 y_6 y_4^2 - y_1 y_7 y_4^2
  + 2 y_2 y_5^2 y_4 + 2 y_3^2 y_6 y_4 + 2 y_1 y_5 y_6 y_4 \nonumber \\*
& & +\ 2 y_2 y_3 y_7 y_4 - y_1 y_5^3 + y_3^2 y_5^2 + y_2^2 y_6^2
  - y_1 y_3 y_6^2 - 2 y_2 y_3 y_5 y_6 - y_3^3 y_7 \nonumber \\*
& & -\ y_2^2 y_5 y_7 + y_1 y_3 y_5 y_7 \nonumber \\
& B=\!\!\!\! & y_8 y_3^3 - 2 y_5 y_6 y_3^2 - y_4 y_7 y_3^2
  + 2 y_4 y_5^2 y_3 + y_2 y_6^2 y_3 + y_4^2 y_6 y_3 + y_2 y_5 y_7 y_3
  \nonumber \\*
& & +\ y_1 y_6 y_7 y_3 - 2 y_2 y_4 y_8 y_3
  - y_1 y_5 y_8 y_3 - y_2 y_5^3 - y_1 y_4 y_6^2 - y_4^3 y_5 \nonumber \\*
& & +\ y_1 y_5^2 y_6 + y_2 y_4^2 y_7 - y_1 y_4 y_5 y_7 - y_2^2 y_6 y_7
  + y_1 y_4^2 y_8 + y_2^2 y_5 y_8 \nonumber \\
& C=\!\!\!\! & - y_6 y_4^3 + y_5^2 y_4^2 + y_3 y_7 y_4^2 + y_2 y_8 y_4^2
  + y_2 y_6^2 y_4 - 3 y_2 y_5 y_7 y_4 + y_1 y_6 y_7 y_4 \nonumber \\*
& & -\ y_3^2 y_8 y_4 - y_1 y_5 y_8 y_4 - y_3 y_5^3 - y_1 y_5 y_6^2
  + y_2^2 y_7^2 - y_1 y_3 y_7^2 + y_2 y_5^2 y_6 \nonumber \\*
& & +\ y_1 y_5^2 y_7 + y_3^2 y_5 y_7 - y_2 y_3 y_6 y_7 + y_2 y_3 y_5 y_8
  - y_2^2 y_6 y_8 + y_1 y_3 y_6 y_8 \nonumber \\
& D=\!\!\!\! & - y_7 y_4^3 + 2 y_5 y_6 y_4^2 + y_3 y_8 y_4^2 - y_5^3 y_4
  - 2 y_3 y_6^2 y_4 + y_1 y_7^2 y_4 + y_2 y_6 y_7 y_4 \nonumber \\*
& & -\ y_2 y_5 y_8 y_4 - y_1 y_6 y_8 y_4 + y_1 y_6^3 - y_2 y_5 y_6^2
  - y_2 y_3 y_7^2 + y_3 y_5^2 y_6 + y_2 y_5^2 y_7 \nonumber \\*
& & +\ y_3^2 y_6 y_7 - 2 y_1 y_5 y_6 y_7 + y_1 y_5^2 y_8 - y_3^2 y_5 y_8
  + y_2 y_3 y_6 y_8 \nonumber \\
& E=\!\!\!\! & y_5^4 - 3 y_4 y_6 y_5^2 - 2 y_3 y_7 y_5^2 - y_2 y_8 y_5^2
  + 2 y_3 y_6^2 y_5 + 2 y_4^2 y_7 y_5 + 2 y_2 y_6 y_7 y_5 \nonumber \\*
& & +\ 2 y_3 y_4 y_8 y_5 - y_2 y_6^3 + y_4^2 y_6^2 + y_3^2 y_7^2
  - y_2 y_4 y_7^2 - 2 y_3 y_4 y_6 y_7 - y_4^3 y_8 \nonumber \\*
& & -\ y_3^2 y_6 y_8 + y_2 y_4 y_6 y_8 \nonumber
\end{eqnarray}
As before, the coefficients $A$ though $E$ are minors of a Hankel matrix of
$y_n$'s, so this equation can be written:
\begin{equation}
\label{eq:V1x8x4determinant}
\left\vert\begin{BMAT}(b){cccc.c}{ccccc}
y_1 & y_2 & y_3 & y_4 & 1     \\
y_2 & y_3 & y_4 & y_5 & x_1   \\
y_3 & y_4 & y_5 & y_6 & x_1^2 \\
y_4 & y_5 & y_6 & y_7 & x_1^3 \\
y_5 & y_6 & y_7 & y_8 & x_1^4
\end{BMAT}\right\vert = 0.
\end{equation}
The quartic equation can be solved using the method of Ferrari
\cite{Cardano:1545}. 

\section{\label{sec:KMN-examples}Solution for examples of $(K,M,N)$}

\subsection{\label{sub:KMN=223}Solution for $(K,M,N) = (2,2,3)$}

The nonlinear equations to solve has a block structure:
\begin{equation}
\left[\begin{array}{c}
  y_{11} \\ y_{12} \\ y_{13} \\ \hline y_{21} \\ y_{22} \\ y_{23}
\end{array}\right] = \left[\begin{array}{cc|cc}
1     & 1     &       &       \\
x_1   & x_2   &       &       \\
x_1^2 & x_2^2 &       &       \\
\hline
      &       & 1     & 1     \\
      &       & x_1   & x_2   \\
      &       & x_1^2 & x_2^2
\end{array}\right] \left[\begin{array}{c}
a_{11} \\ a_{12} \\ \hline a_{21} \\ a_{22}
\end{array}\right]
\end{equation}
Here, the indices for $y_{kn}$, $x_m$ and $a_{km}$ are in the ranges $1 \le k
\le K$, $1 \le m \le M$ and $1 \le n \le N$. To reduce the system we extend
Eq.~(\ref{eq:BP_prefactor}) to block form with $K$ identical blocks $L_m(x)$
on the diagonal. The reduced equations are
\begin{eqnarray}
\label{eq:V2x2x3eq1}
y_{k1} & = & a_{k1} + a_{k2} \\
\label{eq:V2x2x3eq2}
\alpha_{k2} & = & \left( x_1 - x_2 \right) a_{k2} \\
\label{eq:V2x2x3eq3}
\beta_{k3} & = & 0 \\
&& \qquad ( 1 \le k \le 2 ) \nonumber
\end{eqnarray}
where we have added an additional index $k$ to the auxiliary quantities
defined in Eqs.~(\ref{eq:alpha}) and (\ref{eq:beta}). Substituting for
$\beta_{k3}$ in Eqs.~(\ref{eq:V2x2x3eq3}) and eliminating $x_2$ gives the
equation:
\begin{equation}
\label{eq:V2x2x3reduction1}
\left\vert \begin{array}{cc}
  \alpha_{12} & \alpha_{22} \\
  \alpha_{13} & \alpha_{23}
\end{array} \right\vert = 0,
\end{equation}
where we've written the equation as a minor of some matrix, following our
experience in Appendix~\ref{sec:Meff_m234}, yet whose structure is not yet
clear. Substituting for $\alpha_{kn}$ gives a quadratic equation in $x_1$ in
determinant form:
\begin{equation}
\left\vert\begin{BMAT}(b){c.c.c}{ccc}
\label{eq:V2x2x3determinant}
y_{11} & y_{21} & 1     \\
y_{12} & y_{22} & x_1   \\
y_{13} & y_{23} & x_1^2
\end{BMAT}\right\vert = 0 .
\end{equation}

\subsection{\label{sub:KMN=246}Solution for $(K,M,N) = (2,4,6)$}

The reduced system of equations for $K=2$ correlation functions measured on
$N=6$ equally spaced time slices to be solved to extract model parameters for
$M=4$ states is:
\begin{eqnarray}
\label{eq:V2x4x6eq1}
y_{k1}      & = & a_{k1} + a_{k2} + a_{k3} + a_{k4}                         \\
\label{eq:V2x4x6eq2}
\alpha_{k2} & = &   a_{k2} \left( x_1 - x_2 \right)
                  + a_{k3} \left( x_1 - x_3 \right)
                  + a_{k4} \left( x_1 - x_4 \right)                         \\
\label{eq:V2x4x6eq3}
\beta_{k3}  & = &   a_{k3} \left( x_1 - x_3 \right) \left( x_2 - x_3 \right)
                  + a_{k4} \left( x_1 - x_4 \right) \left( x_2 - x_4 \right)\\
\label{eq:V2x4x6eq4}
\gamma_{k4} & = & a_{k4} \left( x_1 - x_4 \right) \left( x_2 - x_4 \right)
                         \left( x_3 - x_4 \right)                           \\
\label{eq:V2x4x6eq5}
\delta_{k5} & = & 0                                                         \\
\label{eq:V2x4x6eq6}
\delta_{k6} & = & 0                                                         \\
&& \qquad ( 1 \le k \le 2) \nonumber
\end{eqnarray}
Substituting for $\delta_{kn}$ and eliminating $x_4$ gives a set of equations
in $\gamma_{kn}$:
\begin{eqnarray}
\left| \begin{array}{cc}
  \gamma_{14} & \gamma_{24} \\
  \gamma_{15} & \gamma_{25} \\
\end{array} \right| = 0 ,      &
\left| \begin{array}{cc}
  \gamma_{14} & \gamma_{25} \\
  \gamma_{15} & \gamma_{26} \\
\end{array} \right| = 0 ,      &
\left| \begin{array}{cc}
  \gamma_{15} & \gamma_{24} \\
  \gamma_{16} & \gamma_{25} \\
\end{array} \right| = 0 ,      \\
\left| \begin{array}{cc}
  \gamma_{15} & \gamma_{25} \\
  \gamma_{16} & \gamma_{26} \\
\end{array} \right| = 0 ,      &
\left| \begin{array}{cc}
  \gamma_{14} & \gamma_{15} \\
  \gamma_{15} & \gamma_{16} \\
\end{array} \right| = 0 ,      &
\left| \begin{array}{cc}
  \gamma_{24} & \gamma_{25} \\
  \gamma_{25} & \gamma_{26} \\
\end{array} \right| = 0 .    \nonumber
\end{eqnarray}
Substituting for $\gamma_{kn}$ and eliminating $x_3$ gives a set of equations
in $\beta_{kn}$:
\begin{equation}
\left|\begin{array}{ccc}
  \beta_{13} & \beta_{14} & \beta_{23} \\
  \beta_{14} & \beta_{15} & \beta_{24} \\
  \beta_{15} & \beta_{16} & \beta_{25} \\
\end{array}\right| = 0, \quad
\left|\begin{array}{ccc}
  \beta_{13} & \beta_{14} & \beta_{24} \\
  \beta_{14} & \beta_{15} & \beta_{25} \\
  \beta_{15} & \beta_{16} & \beta_{26} \\
\end{array}\right| = 0, \quad
\left|\begin{array}{ccc}
  \beta_{13} & \beta_{23} & \beta_{24} \\
  \beta_{14} & \beta_{24} & \beta_{25} \\
  \beta_{15} & \beta_{25} & \beta_{26} \\
\end{array}\right| = 0, \quad
\left|\begin{array}{ccc}
  \beta_{14} & \beta_{23} & \beta_{24} \\
  \beta_{15} & \beta_{24} & \beta_{25} \\
  \beta_{16} & \beta_{25} & \beta_{26} \\
\end{array}\right| = 0 .
\end{equation}
Substituting for $\beta_{kn}$ and eliminating $x_2$ gives an equation in
$\alpha_{kn}$:
\begin{equation}
  \left| \begin{array}{cccc}
    \alpha_{12} & \alpha_{13} & \alpha_{22} & \alpha_{23} \\
    \alpha_{13} & \alpha_{14} & \alpha_{23} & \alpha_{24} \\
    \alpha_{14} & \alpha_{15} & \alpha_{24} & \alpha_{25} \\
    \alpha_{15} & \alpha_{16} & \alpha_{25} & \alpha_{26} \\
  \end{array} \right| = 0
\end{equation}
Substituting for $\alpha_{kn}$ gives a quartic equation in $x_1$:
\begin{equation}
  \label{eq:V2x4x6determinant}
  \left| \begin{BMAT}(b){cc.cc.c}{ccccc}
    y_{11} & y_{12} & y_{21} & y_{22} & 1     \\
    y_{12} & y_{13} & y_{22} & y_{23} & x_1   \\
    y_{13} & y_{14} & y_{23} & y_{24} & x_1^2 \\
    y_{14} & y_{15} & y_{24} & y_{25} & x_1^3 \\
    y_{15} & y_{16} & y_{25} & y_{26} & x_1^4
  \end{BMAT}\right| = 0 .
\end{equation}
%

\subsection{\label{sub:KMN=334}Solution for (K,M,N) = (3,3,4)}

The reduced system of equations for $K=3$ correlation functions measured on
$N=4$ equally spaced time slices to be solved to extract model parameters for
$M=3$ states is:
\begin{eqnarray}
\label{eq:V3x3x4eq1}
y_{k1}      & = & a_{k1} + a_{k2} + a_{k3}                                 \\
\label{eq:V3x3x4eq2}
\alpha_{k2} & = &   a_{k2} \left( x_1 - x_2 \right)
                  + a_{k3} \left( x_1 - x_3 \right)                        \\
\label{eq:V3x3x4eq3}
\beta_{k3}  & = & a_{k3} \left( x_1 - x_3 \right) \left( x_2 - x_3 \right) \\
\label{eq:V3x3x4eq4}
\gamma_{k4} & = & 0                                                        \\
&& \qquad ( 1 \le k \le 3 ) \nonumber
\end{eqnarray}
Substituting for $\gamma_{k4}$ and eliminating $x_3$ gives a set of equations
in $\beta_{kn}$:
\begin{equation}
  \left| \begin{array}{cc}
    \beta_{13} & \beta_{23} \\
    \beta_{14} & \beta_{24} \\
  \end{array} \right| = 0 , \quad
  \left| \begin{array}{cc}
    \beta_{13} & \beta_{33} \\
    \beta_{14} & \beta_{34} \\
  \end{array} \right| = 0 , \quad
  \left| \begin{array}{cc}
    \beta_{23} & \beta_{33} \\
    \beta_{24} & \beta_{34} \\
  \end{array} \right| = 0 .
\end{equation}
Substituting for $\beta_{kn}$ and eliminating $x_2$ gives an equation in
$\alpha_{kn}$:
\begin{equation}
  \left| \begin{array}{ccc}
    \alpha_{12} & \alpha_{22} & \alpha_{32} \\
    \alpha_{13} & \alpha_{23} & \alpha_{33} \\
    \alpha_{14} & \alpha_{24} & \alpha_{34} \\
  \end{array} \right| = 0 .
\end{equation}
Substituting for $\alpha_{kn}$ gives a cubic equation in $x_1$:
\begin{equation}
  \label{eq:V3x3x4determinant}
  \left| \begin{BMAT}(b){c.c.c.c}{cccc}
    y_{11} & y_{21} & y_{31} & 1     \\
    y_{12} & y_{22} & y_{32} & x_1   \\
    y_{13} & y_{23} & y_{33} & x_1^2 \\
    y_{14} & y_{24} & y_{34} & x_1^3
  \end{BMAT}\right| = 0 .
\end{equation}

\subsection{\label{sub:KMN=445}Solution for $(K,M,N) = (4,4,5)$}

The reduced system of equations for $K=4$ correlation functions measured on
$N=5$ equally spaced time slices to be solved to extract model parameters for
$M=4$ states is:
\begin{eqnarray}
\label{eq:V4x4x5eq1}
y_{k1}      & = & a_{k1} + a_{k2} + a_{k3} + a_{k4}                         \\
\label{eq:V4x4x5eq2}
\alpha_{k2} & = &   a_{k2} \left( x_1 - x_2 \right)
                  + a_{k3} \left( x_1 - x_3 \right)
                  + a_{k4} \left( x_1 - x_4 \right)                         \\
\label{eq:V4x4x5eq3}
\beta_{k3}  & = &  a_{k3} \left( x_1 - x_3 \right) \left( x_2 - x_3 \right)
                 + a_{k4} \left( x_1 - x_4 \right) \left( x_2 - x_4 \right) \\
\label{eq:V4x4x5eq4}
\gamma_{k4} & = & a_{k4} \left( x_1 - x_4 \right) \left( x_2 - x_4 \right)
                  \left( x_3 - x_4 \right)                                  \\
\label{eq:V4x4x5eq5}
\delta_{k5} & = & 0                                                         \\
&& \qquad ( 1 \le k \le 4 ) \nonumber
\end{eqnarray}

The determinant form of the quartic equation in $x_1$ is
\begin{equation}
  \label{eq:V4x4x5determinant}
  \left| \begin{BMAT}(b){c.c.c.c.c}{ccccc}
    y_{11} & y_{21} & y_{31} & y_{41} & 1     \\
    y_{12} & y_{22} & y_{32} & y_{42} & x_1   \\
    y_{13} & y_{23} & y_{33} & y_{43} & x_1^2 \\
    y_{14} & y_{24} & y_{34} & y_{44} & x_1^3 \\
    y_{15} & y_{25} & y_{35} & y_{45} & x_1^4
  \end{BMAT}\right| = 0 .
\end{equation}
%

\bibliography{main}

\end{document}